\documentclass[aps,twocolumn,twoside,floatfix,nofootinbib,a4paper]{revtex4}

\usepackage{graphicx}
\usepackage{marginnote}
\usepackage{bbm}
\usepackage{hyperref}
\usepackage{enumitem,kantlipsum}
\usepackage[usenames,dvipsnames]{color}
\usepackage{amsmath}
\usepackage{amssymb}
\usepackage{amsfonts}
\usepackage{mathrsfs}
\usepackage{bm}
\usepackage[all]{xy}
\usepackage{hyperref}
\usepackage{tikz-cd}

\usepackage{graphicx}
\usepackage{marginnote}
\usepackage{hyperref}

\usepackage{enumitem,kantlipsum}
\usepackage{amsmath}
\usepackage{amssymb}
\usepackage{amsfonts}
\usepackage{mathrsfs}
\usepackage{bm}
\usepackage[all]{xy}
\usepackage{hyperref}
\usepackage{tikz-cd}

\newcommand{\beq}{\begin{equation}}
\newcommand{\eeq}{\end{equation}}
\newcommand{\bea}{\begin{eqnarray}}
\newcommand{\eea}{\end{eqnarray}}
	
\newcommand{\ea}{\end{align*}}

\newcommand{\bma}{\begin{pmatrix}}
\newcommand{\ema}{\end{pmatrix}}

\usepackage[percent]{overpic}
\usepackage{overpic}

\begin{document}

\title{Revisiting neutrino-driven magnetogenesis during stellar core collapse}

\author{Fan Zhang} 
\affiliation{Institute for Frontiers in Astronomy and Astrophysics, Beijing Normal University, Beijing 102206, China}
\affiliation{Gravitational Wave and Cosmology Laboratory, School of Physics and Astronomy, Beijing Normal University, Beijing 100875, China}

\date{\today}

\begin{abstract}
\begin{center}
\begin{minipage}[c]{0.9\textwidth}
The literature has not converged onto a precise depiction of the magnetogenesis process for pulsars, and it is profitable to preliminarily but exhaustively assess the viability of the plethora of alternative proposals, before substantial efforts are invested into simulating them in detail. In this note, we tackle one of them, taking notice of an earlier work that suggests neutrino ponderomotive force could spawn a magnetic field not so far off from pulsar strengths. We reexamine this mechanism with more modern technology, accounting for actual core collapse dynamics, and show that this mechanism is likely less powerful than originally envisioned.  
\end{minipage} 
 \end{center}
  \end{abstract}
\maketitle

\raggedbottom

\section{Introduction}
The generation of strong magnetic fields for pulsars is somewhat of a mystery. The traditional view is that frozen-in field from the progenitor star is able to account for regular pulsar magnetic field strength. However, any frozen-in fields (they may not really be frozen-in even assuming infinite conductivity, whenever neutrinos are present, see e.g., \cite{haas2016neutrino}) are necessarily severely disturbed during CCSNe, so a large scale dipolar field is not a given \citep{burrows2021core}.  Moreover, if a strong magnetic field had always been there, it would couple the rotation of the envelop and core during giant and supergiant phases, slowing down the core rotation to much slower than the neutron stars observed, essentially losing angular momentum to the material blown out into the interstellar medium. 

The recently more popular alternative is to have the fields generated by various dynamo effects (see e.g., \cite{2009IAUS..259...61S} for a summary of various proposals). Beside these mechanisms though, it had also been suggested in the past by \cite{shukla1998intense} that a neutrino ponderomotive force engendered magnetic field could reach an order of $10^8-10^9\text{ G}$. This is tantalizingly close to the level one needs for the pulsars. 
Indeed, one might argue that neutrinos represent too big of an energy reservoir not to tap into. The total magnetic field energy for a regular pulsar is in the order of $10^{40}\text{ erg}$ (see e.g., \cite{2017A&A...598A..88Z}), ten or more orders of magnitude less than the energy available within the neutrino sector, so even an inefficient conversion mechanism is possibly allowed. Furthermore, there is this neutrino trapping phase in the core when density rises to above $\approx 10^{11} \text{ g cm}^{-3}$  (see e.g., \cite{mazurek1974degeneracy,sato1975supernova,1977ApJ...218..815A,burrows2013colloquium}), which happens during the collapse of the core before the bounce. In other words, we have a long lasting soup of very high density neutrinos, electrons and ions, all coupled by weak interactions. Within this soup, neutrinos push against positively and negatively charged particles differently, thus could in principle create currents and subsequently magnetic fields. Furthermore, this soup is not in a static thermo state, but rather hosts bulk charged particle motions, as well as large scale inhomogeneities in neutrino phase space distribution, driven by the inward collapse and the associated spinning-up of the core. As a result, there exists coherent nonlinear neutrino ponderomotive effects that could produce magnetic fields that are global by birth, rather than being initially fragmented into tiny domains, and have to be collated later somehow.  

However, it is worth noting that this very early study by \cite{shukla1998intense} recognizes only the neutrino density gradient contribution to the ponderomotive force and misses the more important neutrino flux contributions. In fact, the gradient-induced force should not produce any magnetic field at all (see the discussion below Eq.~\ref{eq:nupotentials}, and also \cite{brizard2000magnetic}). This problem is not noticed by \cite{shukla1998intense} because it does not compute the field explicitly, but instead simply assumes that it exists, and the field strength is estimated indirectly by requiring a balance between the Lorentz and the ponderomotive forces to achieve a stationary equilibrium state. This strategy has its own fallacies even if the gradient force can bring about magnetism, because during core collapse, gravity and strong forces dominate, and the balancing or not of these subdominant effects have no consequence on whether the underlying collapse dynamics, that drives magnetogenesis, ends. In other words, while local stationary equilibrium calculations are easier to perform, it may not be relevant for the physical reality of the supernova process. With these in mind, it would be important to re-assess the viability of neutrino-driven magnetogenesis with more modern technologies, taking from recent modelling efforts, those ingredients that would allow a computation in the context of actual core collapse dynamics, and taking into account the neutrino flux contribution to the ponderomotive force.
In this note, we demonstrate, via detailed such computations, that unfortunately, the neutrino ponderomotive effect is suppressed to such extremes by the weakness of the weak interactions, that even the extraordinary abundance of neutrinos during supernova fails to overcome it.  

\section{Quantitative assessment}
\subsection{Magnetic field growth rate}
The neutrino-coupled magnetohydrodynamics formalisms have recently become available within the plasma community (see e.g., \cite{silva2000neutrino,brizard2000magnetic,haas2016neutrino}). 
The magnetic field evolution is given in Eq.~5.7 of \cite{brizard2000magnetic}, and the part relevant for us is (we assume Gaussian-cgs units)
\begin{align} \label{eq:dtB}
\frac{\partial {\bf  B}}{\partial t} = \sum_{\sigma} \frac{c q_{\sigma}}{Q^2} \nabla \times \mathcal{F}_{\sigma} + \cdots\,, 
\end{align}
where $\nu$ enumerates neutrino species and $\sigma$ the plasma particles, with  $q_{\sigma}$ their electric charge and $Q^2 \equiv \sum_{\sigma}q^2_{\sigma}$. This expression traces back to the neutrino-induced ponderomotive force experienced by plasma particles, given in a Lorentz force analogue expression (${\bf  v}_{\sigma}$ being plasma particle velocity)
\bea \label{eq:Force}
\mathcal{F}_{\sigma} \equiv \sum_{\nu} G_{\sigma \nu} \left[ -\mathcal{E}_{\nu} + {\bf  v}_{\sigma} \times \mathcal{B}_{\nu} \right]\,,
\eea
with the neutrino-electric and neutrino-magnetic fields being defined by the expressions
\bea \label{eq:nupotentials}
\mathcal{E}_{\nu} \equiv \left( \nabla n_{\nu} + \frac{1}{c} \frac{\partial {\bf  J}_{\nu}}{\partial t} \right)\,, \quad \mathcal{B}_{\nu} \equiv \frac{1}{c}  \nabla \times {\bf  J}_{\nu}\,,
\eea
where the neutrino particle number density $n_{\nu}$ (note the curl of a gradient is zero, so $n_{\nu}$'s contribution vanishes in Eq.~\ref{eq:dtB} and $n_{\nu}$ is not relevant for us) and normalized flux ${\bf  J}_{\nu} \equiv n_{\nu} {\bf v}_{\nu}/c$ serve as the analogue to the scalar electric potential and the vector magnetic potential, respectively. 
The charge in this electromagnetic analogue is   
\begin{align}
G_{\sigma \nu} \equiv \sqrt{2} G_{F} \left[\delta_{e\sigma} \delta_{\nu \nu_e} + \left(I_{\sigma}-2Q_{\sigma} \sin^2 \theta_W\right)\right]\,,
\end{align}
whereby $G_F \approx 1.43523\times 10^{-49} \text{ erg cm}^{3}$ is the Fermi constant, $Q_{\sigma} \equiv q_{\sigma}/e$, $I_{\sigma}$ is weak isospin ($-1/2$ for electron and $1/2$ for proton) and $\theta_W$ is the Weinberg angle. The first term is from charged weak currents (it suffices to include only the first generation of particles), and the rest are from neutral weak currents.  

We will see later that the spin of the collapsing core will contribute to the bulk parts of ${\bf  v}_{\sigma}$ and ${\bf  J}_{\nu}$ (additional thermal randomness are also present but not relevant for large scale magnetic field generation). To incorporate it, it suffices to consider the still simplified situation of axisymmetry. We have then that the vertical (along the rotational axis) component of the growth rate of the magnetic field becomes
\begin{align} \label{eq:dtBz}
\frac{\partial B^z}{\partial t} = & \sum_{\sigma}\sum_{\nu}
-\frac{G_{\sigma \nu} q_{\sigma}}{Q^2 r^2} \bigg\{v_{\sigma}^r  \bigg[r \bigg( \cos\theta  \frac{\partial ^2 J_{\nu}^{\phi}}{\partial r \partial \theta }
+r \sin\theta  \frac{\partial ^2 J_{\nu}^{\phi}}{\partial r^2}
\notag \\ &
+\frac{\partial  J_{\nu}^{\phi}}{\partial r} (2 \sin\theta +\cos\theta \cot\theta )\bigg)
+\cos\theta  \frac{\partial  J_{\nu}^{\phi}}{\partial \theta }\bigg]
\notag \\ &
+ J_{\nu}^{\phi}  \bigg[ \cos\theta  \cot\theta  v_{\sigma}^r 
+r \sin\theta  \frac{\partial v_{\sigma}^r}{\partial r}+\cos\theta  \frac{\partial v_{\sigma}^r}{\partial \theta }\bigg]
\notag \\ &
+r \bigg[r \sin\theta  \frac{\partial  J_{\nu}^{\phi}}{\partial r} \frac{\partial v_{\sigma}^r}{\partial r}
+\cos\theta  \frac{\partial  J_{\nu}^{\phi}}{\partial r} \frac{\partial v_{\sigma}^r}{\partial \theta }
\notag \\ &
+r \sin\theta  \frac{\partial ^2 J_{\nu}^{\phi}}{\partial t \partial r}
+\cos\theta  \frac{\partial ^2 J_{\nu}^{\phi}}{\partial t \partial \theta }
\notag \\ &
+(\sin\theta  +\cos\theta  \cot\theta)  \frac{\partial  J_{\nu}^{\phi}}{\partial t}\bigg]\bigg\}\,.
\end{align}
This component is important as it could not only underpin the dipolar pulsar field, but also help collimate outward exploding material into jets, which is assumed to be responsible for long gamma ray bursts \citep{piran2005physics}. It is in general nonvanishing, thanks to the different behaviour between electrons and ions, ensuring their contributions don't cancel in the $\sum_{\sigma}$ of Eq.~\ref{eq:dtBz}.

To have an intuitive feel of where $\partial_t {B}^{z}$ comes from, we take a closer look at the terms in Eq.~\eqref{eq:dtBz}. First note that all the terms are linear in ${J}^{\phi}_{\nu}$\footnote{This is the component of ${\bf  J}_{\nu}$ in an orthonormal basis aligned with the spherical coordinate system, i.e., the metric for its contraction with other vectors does not contain factors of $r$.} or its derivatives. With it, we would obtain a $\mathcal{B}_{\nu}^z$ by Stokes' theorem, which crosses with the radial component ${v}^r_{\sigma}$ of the infalling plasma, to create an $\mathcal{F}_{\nu}^{\phi}$, which in turn gives rise to a $\partial_t {B}^{z}$, once again by Stokes' theorem. This accounts for the terms linear in $v_{\sigma}^{r}$, while the bottom two lines containing no $v_{\sigma}^{r}$, but instead temporal derivatives, are from $\mathcal{E}_{\nu}$. For the next couple of sections, we turn to obtaining the values of the various quantities appearing in Eq.~\eqref{eq:dtBz}.  

\vspace{5mm}
\subsection{Plasma compression speed \label{sec:Ana}}
Fortunately, to zeroth order (1-D spherically symmetric radial motion only), ${\bf  v}_{\sigma}$ during the entire collapse phase can be obtained from the analytical computations of \cite{kwak2015core} (numerical results such as those displayed in \cite{1977ApJ...218..815A} zooms close in on near-bounce epoch only), which gives
\bea \label{eq:AnaVelo}
{\bf v}^{(0)}_{\sigma}(r,t) = \frac{\dot{R}}{R}{\bf r}\,, 
\eea
where $R$ is the overall core radius, $r$ is the radial coordinate and overhead dot denotes time derivative. This expression is supplemented by 
\bea \label{eq:AnaR}
R= \frac{1}{2}R_{\rm max} \left(1+\cos \psi \right)\,, \quad
\dot{R} =-R_{\rm max}\frac{k\sin\psi}{1+\cos\psi}\,, 
\eea
where 
\bea
k = \sqrt{\frac{4(4-3\gamma)GM}{R^3_{\rm max}}}\,,
\eea
with $G$ being the gravitational constant, $M$ the core mass, and $\gamma$ the specific heat ratio or exponent in the polytrope equation of state at the center of the core (the coefficient in the polytrope on the other hand, is set so that the effective potential of the radial collapse vanishes initially).
In addition, the computations in \cite{kwak2015core} also yield a nucleon mass density 
\bea \label{eq:AnaDen}
\rho_{\rm nuc}(r,t) =\frac{15M}{8\pi R^3}\left(1-\frac{r^2}{R^2}\right)\,,
\eea
which will be useful when we lift data from numerical simulations. Specifically, for economy of space, the simulations usually tabulate results only for some selected snapshot time close to core bounce, but nevertheless those radial shells at other times have been noted \citep{1977ApJ...218..815A} to share physical attributes\footnote{Note though ${\bf v}_{\sigma}$ for common density shells at different times would in general differ, and we must use Eq.~\eqref{eq:AnaVelo}, and not the mapped-onto snapshot-specific numerical in-fall velocity.} with one particular shell at the displayed time, namely the one that is similar in $\rho_{\rm nuc}$. Consequently, the analytical expression \eqref{eq:AnaDen}, available for all times and radius during collapse, allows us to locate the relevant numerical neutrino energy flux data that is sensitive to matter density.  

The numerical study we will rely on for neutrino information is \cite{1977ApJ...218..815A}\footnote{This work predates many new advances in supernova and particle theories, e.g., its Weinberg angle is not quite right. However, it studies neutrino trapping for the pre-bounce stage, thus contains details regarding neutrinos during the epoch relevant for us, while more modern simulations concentrate more on the post-bounce explosion, so we had not been able to find the desired details from them.}, and we choose the parameters in the analytical expressions \eqref{eq:AnaVelo} through \eqref{eq:AnaDen} in such a way that agreements with the numerical results are optimized. Specifically, we take zone 27 in \cite{1977ApJ...218..815A}, which is the furthest out co-moving mass shell within the outer core region that is included in their tables, to be the surface of the core, and consequently take the enclosed mass of $M=2.53\times 10^{33} \text{ g}$ ($\approx 1.27 \text{M}_{\odot}$) to be the core mass. The temporal parameter $\psi$ appearing in Eq.~\eqref{eq:AnaR} is related to regular time by
\bea \label{eq:psi}
t= \frac{\psi + \sin\psi}{2 k}\,,
\eea
and starts from $0$ when collapse begins at a core radius of $R_{\rm max} \approx 3000 \text{ km}$. It never reaches $\pi$ where $R=0$, but instead stops at some value $\psi_{\rm max}$ when $R=R_{\rm min} \approx 41.58 \text{ km}$. At this point, a near nuclear density of $\rho_{\rm nuc}=2.1\times 10^{13} \text{ g cm}^{-3}$ (matching the near-bounce central density in \cite{1977ApJ...218..815A}) is reached at the center, at which point we assume that core collapse stops. This termination condition is imposed by hand, rather than automatically, according to an evolution of the equation of state from first principles, because the detailed microphysics is as yet unclear. We also take $\gamma = 1.33038$ during collapse, which produces a final $v_{\sigma}^r = -1.19\times 10^9 \text{ cm s}^{-1}$ at the core surface, matching the result in \cite{1977ApJ...218..815A}. This rather stiff equation of state choice results in a long total collapse time of $t_{\rm max} \approx 3.34 \text{ s}$.

\begin{figure}
\begin{overpic}[width=0.9\columnwidth]{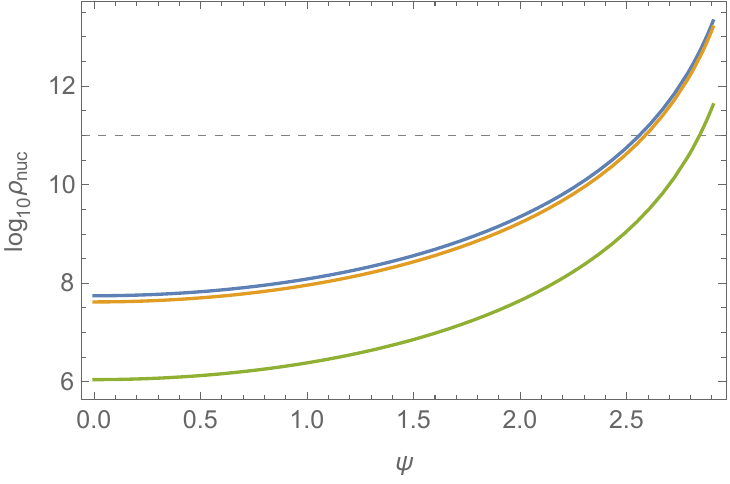}
\end{overpic}
\caption{The analytical $\rho_{\rm nuc}$ as a function of the temporal variable $\psi$ (Eq.~\ref{eq:AnaDen}). The blue, orange and green curves correspond to the center of the core, half way to surface (i.e., at $R/2$), and $99\%$ to surface. The horizontal dashed line marks out $\log_{10}\rho_{\rm nuc} = 11$. 
}
\label{fig:rhovstime}
\end{figure}

\begin{figure}
\begin{overpic}[width=0.9\columnwidth]{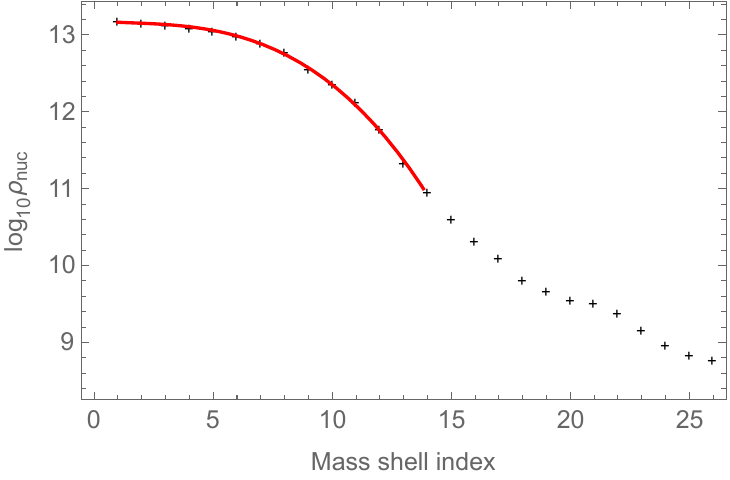}
\end{overpic}
\caption{The numerical $\rho_{\rm nuc}$ as seen in the simulation of \cite{1977ApJ...218..815A}, as a function of mass shell index, a fluid frame radius measure. The data (crosses) are taken from their Tb.~1. We also display, as the red curve, a fit to the numerical data. Only the density region adequate for neutrino trapping is fitted. 
}
\label{fig:rhonucnum}
\end{figure}

To find the numerical data corresponding to any $(\psi,r)$ combination during collapse, we first divide the $\psi_{\rm max}$ interval into $150$ bins, and then for each bin, we further divide the radius range $(0,R(\psi)]$ into another $150$ sub-bins. This way, we obtain a temporal-spatial grid, and for each bin within, we can compute the analytical $\rho_{\rm nuc}(r,\psi)$, which is to be compared with the numerical values tabulated in \cite{1977ApJ...218..815A} for the mass shells at near-bounce time, in order to figure out which of these shells share the most similar environment for neutrinos to our chosen bin. For our purpose, we only need to keep those bins that have a $\rho_{\rm nuc} > 10^{11} \text{ g cm}^{-3}$, or else the neutrinos won't be trapped within. Fig.~\ref{fig:rhovstime} displays the density versus time profile. We see that even for the center of the core, the trapping density is achieved only near the very end of the collapse, at time $\psi=2.5593$ corresponding to $t=3.3058 \text{ s}$. 
We also show in Fig.~\ref{fig:rhonucnum} the $\rho_{\rm nuc}$ distribution just before bounce, as a function of the mass shell index, for which the data is taken from Tb.~1 of \cite{1977ApJ...218..815A}. By comparing with this data, we can assign each temporal-spatial bin a mass shell index number, which serves as the radial coordinate in the fluid frame utilized by the numerical study. 

\vspace{5mm}

\subsection{Neutrino azimuthal flux}
We now turn to the neutrinos. They are trapped in the core towards the end of the collapsing phase, after density reaches $10^{11}\text{g cm}^{-3}$, and must be co-spinning with the plasma material, against which it is constantly interacting and thus strongly coupled.  
The in-falling material is being spun up by the conservation of angular momentum during collapse, so an appreciable bulk $J_{\nu}^{\phi}$ is to be expected. 
To evaluate this flux, we look to the detailed numerical study carried out on the collapse phase by \cite{1977ApJ...218..815A}\footnote{Note the collapse speed $v^r_{\sigma}$ is comparable to the speed of sound at around $10^9 \text{ cm s}^{-1}$ (see fig.~5 of \cite{1977ApJ...218..815A}), not speed of light, thus $v^r_{\sigma}/c \sim 1/10$ is a small expansion parameter, and the results of \cite{1977ApJ...218..815A}, enlisting the radiation transport technology of \cite{1972ApJ...178..779C}, are only valid to first order in $v^r_{\sigma}/c$.}, which provides information on neutrinos in the fluid frame (we denote quantities in this frame with a hat) that compresses in synchrony with the collapsing core, i.e., the radius coordinate in the fluid frame is replaced by the total mass enclosed $\hat{\mathcal{M}}_r$, satisfying
\bea \label{eq:FluidToInertial}
\frac{\partial \hat{\mathcal{M}}_r}{\partial r} = 4\pi r^2 \rho_{\rm nuc}\,, \quad \frac{\partial \hat{\mathcal{M}_r}}{\partial t}= v_{\sigma}^r 4\pi r^2 \rho_{\rm nuc}\,.   
\eea

\begin{figure}
\begin{overpic}[width=0.9\columnwidth]{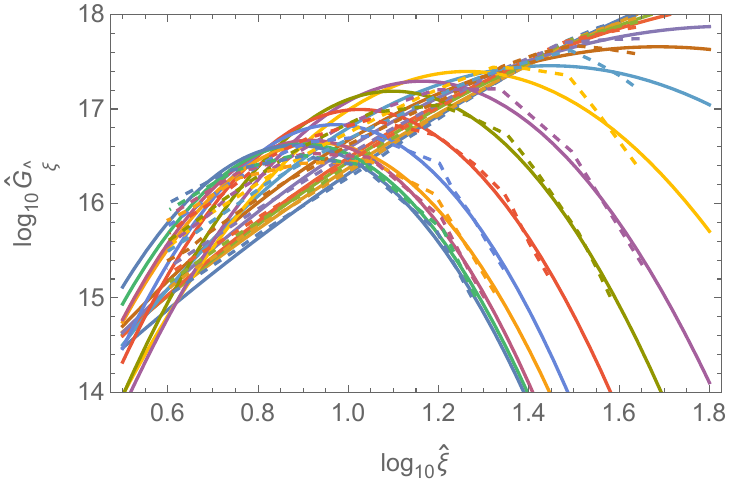}
\end{overpic}
\caption{The numerical $\hat{G}_{\hat{\xi}}$ as seen in the simulation of \cite{1977ApJ...218..815A}, where the original numerical data (from their Fig.~11) is presented as dashed lines, and the quadratic fitting curves are smooth, sharing the same color. Each curve corresponds to a mass shell, with the ones topping out at lower $\hat{G}_{\hat{\xi}}$ values being further out from the center.  
}
\label{fig:GPlot}
\end{figure}

The most important data we need is the neutrino energy per nucleon $\hat{G}_{\hat{\xi}}$, as a function of individual neutrino energy $\hat{\xi}$, displayed in Fig.~11 of \cite{1977ApJ...218..815A}. To acquire it, we need to carry out a fitting step because \cite{1977ApJ...218..815A} truncates the $\hat{G}_{\hat{\xi}}$ data at some neutrino energy and thus does not cover the full range of what we need, so the fitting is used to extrapolate as well as interpolate. Ideally, the numerical $\hat{G}_{\hat{\xi}}$ would be fitted to the expression ($\hat{a}$ equals neutrino chemical potential in equilibrium, $\hat{G}_{\hat{\xi}}^{\text{max}}$ is the limiting value at complete neutrino degeneracy, and $\hat{T}$ is temperature)
\bea \label{eq:Thermo}
\hat{\tilde{G}}_{\hat{\xi}} = \frac{\hat{G}_{\hat{\xi}}^{\text{max}}}{1+ \exp\left[(\hat{\xi}-\hat{a})/k\hat{T}\right]}\,,
\eea
which would be valid if neutrinos are in thermo-equilibrium. Not all strata within the collapsing core are in proper thermo-equilibrium though, and we haven't been able to control the shape of the fitting curves without artificially imposing some $\hat{G}_{\hat{\xi}}^{\text{max}}$ value for the innermost shells, for which the top of the curve isn't displayed by \cite{1977ApJ...218..815A}. The numerical data not being very smooth presents an additional hurdle to this approach. As a result, we resort to simply fitting the log-log plot of $\hat{{G}}_{\hat{\xi}}$ to a quadratic function (this ensures that each curve tops out at some $\hat{G}_{\hat{\xi}}^{\text{max}}$ value rather than keep rising or oscillating wildly; we also weigh larger $\hat{\xi}$ points more heavily, with weighting proportional to $(\log_{10}\hat{\xi})^4$). The fitting result is displayed in Fig.~\ref{fig:GPlot}.  

\begin{figure}
\begin{overpic}[width=0.9\columnwidth]{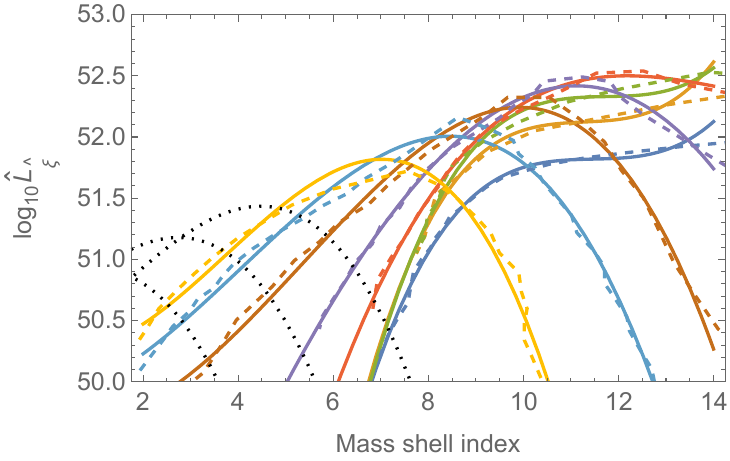}
\end{overpic}
\caption{The numerical $\hat{\mathbb{L}}_{\hat{\xi}}$ as seen in the simulation of \cite{1977ApJ...218..815A}, where the original numerical data (from their Fig.~9) is presented as dashed lines, and the cubic fitting curves are smooth, sharing the same color. Each curve corresponds to a different neutrino energy $\hat{\xi}$, with the higher energy ones generally residing more to the left, closer to the center of the collapsing core. The black dotted curves are extrapolated ones for energies higher than those tabulated by the numerical study. Only neutrino trapping regions (up to mass shell index $14$) are displayed. 
}
\label{fig:LPlot}
\end{figure}

There is also the neutrino luminosity $\hat{\mathbb{L}}_{\hat{\xi}}$ in the fluid frame that contributes to the neutrino flux (although it doesn't contribute directly to our $\partial_t B^z$ estimate under the rigid rotation assumption, we include its computations to demonstrate what can be achieved if we allow differential rotation in the core). This luminosity can be quite appreciable for less than perfectly trapping regions (which really are all the mass shells if we don't confine our consideration to only the highest energy neutrinos), as per Fig.~9 in \cite{1977ApJ...218..815A}, from which we read off our data. We also carry out a fit to the numerical $\hat{\mathbb{L}}_{\hat{\xi}}$ for convenience. Since we only need data up to the $14$th mass shell (neutrinos are not trapped outside of it) at the boundary of the inner core, and the numerical data are rather smooth in this regime, we can simply carry out a cubic polynomial fit, the result of which is shown in Fig.~\ref{fig:LPlot}.

\begin{figure}[b]
\begin{overpic}[width=0.9\columnwidth]{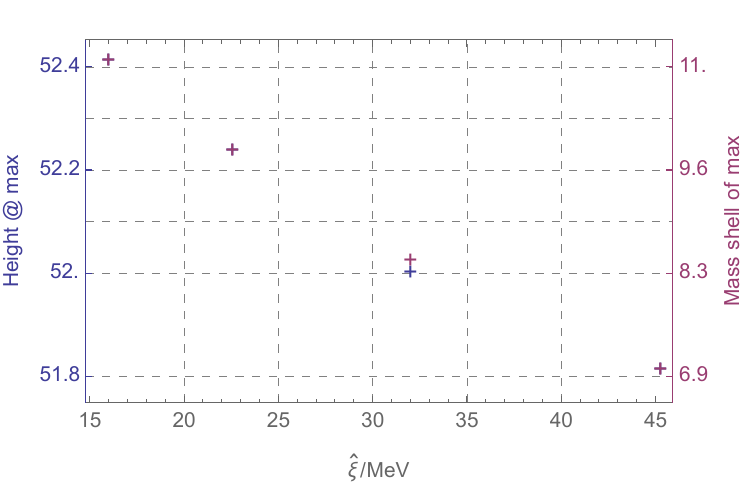}
\end{overpic}
\caption{The $\hat{\xi}$ dependence of the curves in Fig.~\ref{fig:LPlot}. The points ranging from the lowest to the highest energies correspond to the purple, brown, turquoise, and yellow curves, respectively. The blue markers (left axis) show the $\log_{10}\hat{\mathbb{L}}_{\hat{\xi}}$ value achieved at the peaks of these curves in Fig.~\ref{fig:LPlot}, while the magenta markers (right axis) display the mass shell index of the peaks. 
}
\label{fig:LPlotEnergy}
\end{figure}

Figs.~\ref{fig:GPlot} and \ref{fig:LPlot} tabulate the neutrino flux information at near bounce time, and we have to map them onto the bins in our temporal-spatial grid. Before that though, we must note that all these quantities are monochromatic (but not spectral densities, they are the integrated total across $\hat{\xi}$ intervals), so we also bin neutrino energy $\hat{\xi}$ into $100$ intervals running from $3 \text{ MeV}$ to $300 \text{ MeV}$. Fig.~\ref{fig:GPlot} shows that mean energy of escaped neutrinos is $8 \text{ MeV}$, so the trapped neutrinos would typically have higher energies. While the quadratic fit we carried out for Fig.~\ref{fig:GPlot}, and indeed the trend of the underlying numerical data, indicates that there could be extremely energetic neutrinos near the very center of the core, we limit the upper bound for neutrino energy at the low value of $300 \text{ MeV}$ (literature estimates that the \emph{average} neutrino energy in the inner core is $100-300 \text{ MeV}$, see e.g., \cite{burrows2013colloquium}). We do so to avoid extrapolating too aggressively and picking up over-optimism resulting from the naive quadratic fit to $\hat{G}_{\hat{\xi}}$, which appears to decline slower than numerical data at higher energies (see Fig.~\ref{fig:GPlot}). In any case, some experimentation shows that increasing this energy cap hardly changes the result, which is to be expected as the extremely high energy neutrinos only appear at the very center of the core (further inwards from the sampled radius for $\partial_t B^z$ evaluation in Sec.~\ref{sec:BRes} below) at very close to bounce. 

Even with our rather low energy ceiling, the data for $\hat{\mathbb{L}}_{\hat{\xi}}$ needs to be extended to higher $\hat{\xi}$ values than available from numerical data (goes up to around $45 \text{ MeV}$), for which neutrino trapping is more effective because the neutrino-plasma interaction cross sections increase with energy, so $\hat{\mathbb{L}}_{\hat{\xi}}$ would be lower. This trend is already apparent in Fig.~\ref{fig:LPlot} and we distill it further in Fig.~\ref{fig:LPlotEnergy}. We conform to this trend by extrapolating the parameters of the cubic fit against energy. We take the purple, brown, turquoise and yellow curves in Fig.~\ref{fig:LPlot}, corresponding to the higher end energies of $16 \text{ MeV}$, $22.6 \text{ MeV}$, $32 \text{ MeV}$, $45.3 \text{ MeV}$, and linearly (over-fitting with higher order polynomials may result in rising luminosities at very high energies, contradicting physical expectations) extrapolate their peak location and peak height to higher energies. The $\hat{\xi} = 45.3 \text{ MeV}$ curve is then shifted and scaled accordingly to mock up the luminosity curves at these higher extrapolated energies, some of which ($62.5 \text{ MeV}$, $75 \text{ MeV}$ and $87.5 \text{ MeV}$) are shown as black dotted curves in Fig.~\ref{fig:LPlot}. Since the luminosity of higher energy neutrinos decline quite rapidly, they don't contribute significantly to our final result, and our crude extrapolation procedure should suffice. 

\begin{figure}[t]
\centering
\begin{overpic}[width=0.79\columnwidth]{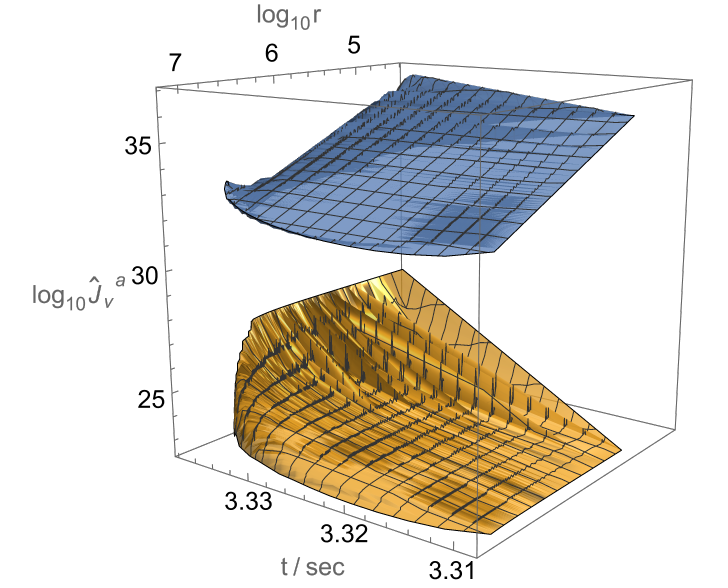}
\put(5,10){(a)}
\end{overpic}
\begin{overpic}[width=0.79\columnwidth]{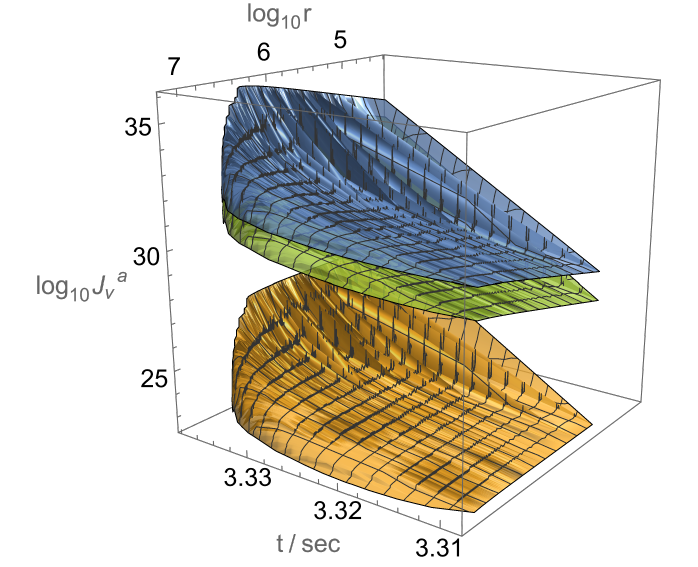}
\put(5,10){(b)}
\end{overpic}
\caption{(a) The fluid frame components $\hat{J}_{\nu}^t$ (bottom yellow surface) and $\hat{J}_{\nu}^r$ (top blue surface) as functions of time and radius (in units of cm before taking logarithm) during core collapse (note these components differ by a factor of $c$ in their definitions, thus are not strictly comparable). (b) The observer frame components ${J}_{\nu}^t$ (bottom yellow surface), ${J}_{\nu}^r$ (top blue surface) and ${J}_{\nu}^{\phi}$ (middle green surface), as functions of time and radius during core collapse. This figure is plotted at $\theta=\pi/4$. 
}
\label{fig:chi}
\end{figure}

We are now ready to fill our spatial-temporal grid with the flux data. For each bin, we find its corresponding sibling mass shell at near bounce time, as per prescription given at the end of Sec.~\ref{sec:Ana}, and for each of the energy bin we read off the $\hat{G}_{\hat{\xi}}$ and $\hat{\mathbb{L}}_{\hat{\xi}}$ values from the fitting functions obtained above.
Now that the energy density and luminosity become available throughout the collapse duration and across the core, we can divide the fluid frame monochromatic energy flux 4-vector 
\begin{align}\label{eq:fluxdensity}
\hat{\zeta}^{t}_{\hat{\xi}} = \rho_{\rm nuc} \hat{G}_{\hat{\xi}} \,, \quad 
\hat{\zeta}^{\mathcal{M}_r}_{\hat{\xi}} =  \frac{\hat{\mathbb{L}}_{\hat{\xi}}}{4\pi r^2}\,, \quad
\hat{\zeta}^{\theta}_{\hat{\xi}} = 0 = \hat{\zeta}^{\phi}_{\hat{\xi}}\,,
\end{align}
by $\hat{\xi}$ (after converting to ergs) to arrive at a monochromatic particle number flux, which is subsequently summed over all energy bins\footnote{Note, we do not know the bin width for the $\hat{G}_{\hat{\xi}}$ displayed by \cite{1977ApJ...218..815A} Fig.~11, and assume it is the same as our sampling bins. This could contribute up to an order of magnitude uncertainty in the final $\partial_t B^z$ estimate.} to yield a proper number flux four vector $c \hat{J}_{\nu}^a$ (Latin alphabet used here for spacetime index; recall that the $\hat{J}_{\nu}^a$ we need in the end is a normalized flux, with unit $\text{cm}^{-3}$ and the flux obtained here is not yet normalized, thus the $c$ factor) that we can transform with the Jacobians of two consecutive frame changes to take us into the observer frame. The non-vanishing components of the $\hat{J}_{\nu}^a$ vector, over the temporal-spatial grid, is displayed as a 3-D graph in Fig.~\ref{fig:chi} (a).

The two frame transformations firstly take us into the co-rotating but non-collapsing frame via a boost parameterized by $v_{\sigma}^r$ (per Eq.~\ref{eq:FluidToInertial}), to pick up the convective flux, and finally into the observer frame in which the core is spinning. The $\phi$ component of the observer frame expression of the normalized number flux vector is then the $J_{\nu}^{\phi}$ appearing in Eq.~\eqref{eq:dtBz}\footnote{After applying the necessary rescaling to transition from the raw coordinate basis in which the metric depends on $r$, into the orthonormal frame. This step introduces a simple $\theta$ dependence.}, that we are looking for.
To find the spin angular velocity needed for the second transformation, we note that the collapse of the core conserves angular momentum, or in other words, the value of (we are assuming rigid rotation, so $\Omega$ has no $r$ dependence) 
\bea \label{eq:Omega}
\int_{0}^{2\pi}\int_{0}^{\pi}\int_{0}^{R(\psi)} && \Omega(\psi) r^4 \sin(\theta)^3 \rho_{\rm nuc}(r,\psi) dr d\theta d\phi \notag \\
&&= \frac{2}{7} \Omega(\psi) M R_{\rm max}^2 \cos^4\left(\frac{\psi}{2}\right)\, 
\eea
remains constant. Therefore, we arrive at  
\bea
\Omega = \Omega_{\rm final} \frac{\cos^4\left(\frac{\psi_{\rm max}}{2}\right)}{\cos^4\left(\frac{\psi}{2}\right)} \approx \frac{0.012}{\cos^4\left(\frac{\psi}{2}\right)}  \,,
\eea
where we have assumed a newborn proto-neutron star period of $0.1 \text{ s}$. The final observer frame ${J}_{\nu}^a$ vector is displayed in Fig.~\ref{fig:chi} (b). 

\subsection{Magnetic field strength \label{sec:BRes}}
\begin{figure}[t]
\begin{overpic}[width=0.9\columnwidth]{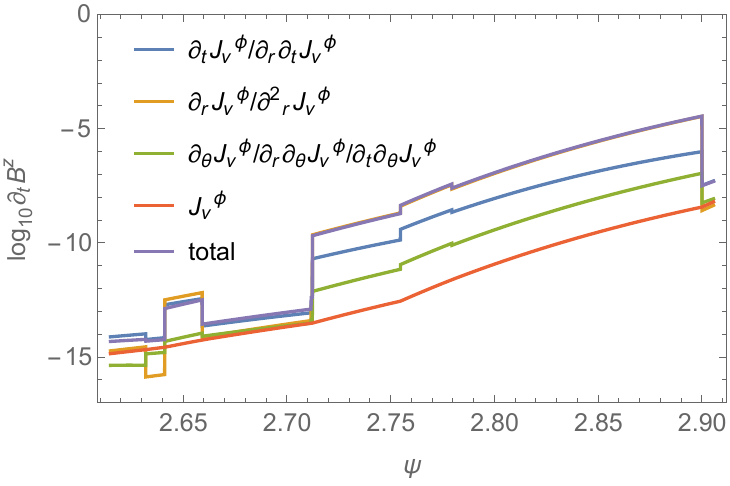}
\end{overpic}
\caption{The $\partial_t B^z$ produced according to the various terms in Eq.~\eqref{eq:dtBz}. Note some terms have opposite signs, but absolute value is taken before applying the logarithm. The minimum and maximum $\psi$ (two ends of the horizontal axis) for which the chosen radius of $r=10 \text{ km}$ is in the neutrino trapping regime correspond to $t_{\rm begin}=3.315 \text{ s}$ and $t_{\rm end}=3.338 \text{ s}$ respectively. The curve labels indicate which terms in Eq.~\eqref{eq:dtBz} are included, e.g., $\partial_t J_{\nu}^{\phi}/\partial_r\partial_t J_{\nu}^{\phi}$ means the sum of contributions from all terms proportional to these two kinds of derivatives on $J_{\nu}^{\phi}$ is being described by the corresponding curve.   
}
\label{fig:dtB}
\end{figure}

We have now acquired all the ingredients that go into Eq.~\eqref{eq:dtBz}, and we can pick a spatial spot and begin evaluating $\partial_t B^z$. As a typical spot inside of the protoneutron star, we pick $r=10\text{ km}$ and $\theta =\pi/4$.
For the plasma species, we sum over electrons and protons in $\sum_{\sigma}$, and we also only consider electron neutrinos (these choices will not change the order of magnitude of the resulting magnetic field). 
With these choices and approximations, we arrive at Fig.~\ref{fig:dtB}, depicting the $\partial_t B^z$ achievable via the neutrino ponderomotive force. 

From Fig.~\ref{fig:chi} (b), we see that there are some ripples on the flux surfaces, which are due to a combination of ruggedness within the original numerical data, and the fact that our data lifting process contains many discrete choices (the rippling increases when we reduce the number of temporal-spatial bins) that jumps between mass shells and fitting curves (e.g., if a desired data point sits between two curves, data is taken from the one that's closest, rather than using a weighted average over all nearby curves). This, and the interpolation procedure employed to obtain the derivatives, leads to some ruggedness in the $\partial_t B^z$ curves in Fig.~\ref{fig:dtB}. 

\subsection{Ponderomotive force strength}
\begin{figure}[t]
\begin{overpic}[width=0.9\columnwidth]{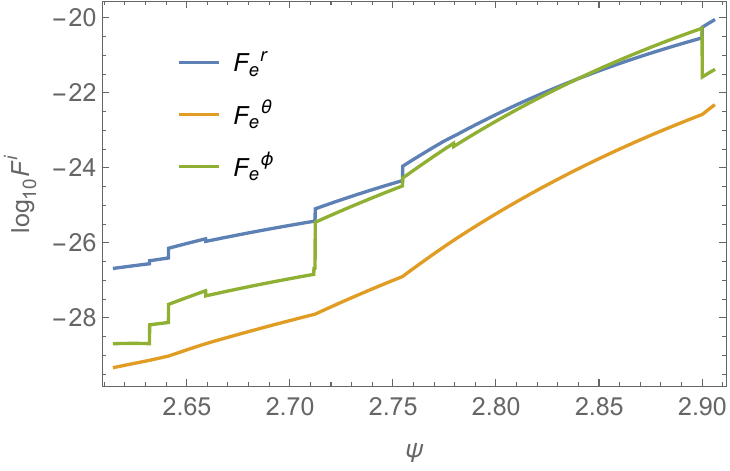}
\end{overpic}
\caption{The ponderomotive force $\mathbb{\bf F}_{e}$ acting on electrons produced according to Eq.~\eqref{eq:ForceCom}. The values are for components expressed in an orthonormal basis aligned with the spherical coordinates. The force acting on protons has the same shape but $\approx 4\%$ of the strength.   
}
\label{fig:Force}
\end{figure}

We can furthermore evaluate the strength of the ponderomotive force during the collapse, in order to help gauge whether collective neutrino-plasma interaction matters at all during supernova. The explicit component form of Eq.~\eqref{eq:Force} within our context is given by 
\begin{align} \label{eq:ForceCom}
\mathcal{F}^r_{\sigma} =& \sum_{\nu}\frac{G_{\sigma \nu}}{c} \left[ \Omega \sin(\theta) \left(J_{\nu}^{\phi} + r\frac{\partial J_{\nu}^{\phi}}{\partial r} \right)- \frac{J_{\nu}^{r}}{\partial t}\right]\,,
\notag \\
\mathcal{F}^{\theta}_{\sigma} =& \sum_{\nu}\frac{G_{\sigma \nu}}{c}\Omega \sin(\theta) \left(\cot(\theta)J_{\nu}^{\phi} + \frac{J_{\nu}^{\phi}}{\partial \theta}\right)\,,
\notag \\
\mathcal{F}^{\phi}_{\sigma} =& \sum_{\nu}-\frac{G_{\sigma \nu}}{cr} \left[ v_{\sigma}^r \left(J_{\nu}^{\phi} + r \frac{\partial J_{\nu}^{\phi}}{\partial r} \right)+r \frac{J_{\nu}^{\phi}}{\partial t}\right]\,,
\end{align}
and their numerical values for the neutrino trapping regime is given in Fig.~\ref{fig:Force}, for which we have adopted the same parameter settings to Fig.~\ref{fig:dtB}.

\section{Discussion and Conclusion}
We have shown, via somewhat detailed computations, that the neutrino ponderomotive force will unlikely produce any appreciable large scale magnetic field. While the parameters chosen in this note are not the most aggressive possible, the suppression coming from the extremely small Fermi constant (a similar ponderomotive process driven by photons may perform better, but the details of photon flux during collapse do not appear to be readily available in literature) would be difficult to overcome by picking more optimistic core collapse physics. 
One thing to note for example, is that although we have computed the radial neutrino number flux in the fluid frame, it does not in fact mix into $J_{\nu}^{\phi}$ for the rigidly rotating core. If there is differential rotation however, it begins to contribute. Nevertheless, even at its full strength of around an order of magnitude larger than the rigid-rotation azimuthal component, it will not be sufficient to raise $\partial_t B^z$ to anything significant. 

Beside arriving at a pessimistic viability assessment for a potential magnetogenesis mechanism, we communicate this null result to also assess the potential gain from introducing more sophisticated neutrino physics into numerical simulations. In this regard, note that to fully account for the neutrinos' impact on other matter, one would ideally simulate detailed interactions at individual particle level, but the associated cost is prohibitive. So we have to resort to broad-stroke coarse-grained approximations, the lowest order of which is an account on energy deposition which does not describe bulk momentum transfer. The next level up includes the ponderomotive forces, which prescribe also some momentum transfer details resulting from collective neutrino-plasma interaction, for which the neutrinos are depicted in the classical fluid style. Our computation in this note suggests that the impact of such first order effects tends to be small during supernova. Specifically, one can compare the force strength as displayed in Fig.~\ref{fig:Force} with the gravitational force experienced by an electron sitting on the surface of the proto-neutron star at the end of collapse $t_{\rm end}$, which is $\approx 10^{-14} \text{ dyn}$. We caution though, our result refers to global scale neutrino distributions, and local turbulent motions involving order unity relative amplitude fluctuations in neutrino flux, occurring at the length scale of millimeters or below, could still generate neutrino ponderomotive forces comparable in strength to gravity. Such a situation may well be relevant for the study of shock revival, where the explosion-aiding turbulence could itself be neutrino driven (see e.g., \cite{burrows2013colloquium}), even via the ponderomotive force itself (see e.g., \cite{silva2000neutrino,haas2016neutrino}).

\begin{acknowledgments}
This work is supported by the National Key Research and Development Program of China grant 2023YFC2205801, and the National Natural Science Foundation of China grants 12433001, 12021003. 
\end{acknowledgments}


\bibliography{supermag.bbl}

\end{document}